\documentclass[aps,pra,twocolumn]{revtex4}%
\usepackage{amssymb}
\usepackage{graphicx}
\usepackage{amsmath}
\usepackage{amsbsy}
\usepackage{cancel}
\usepackage{amsfonts}
\usepackage{amssymb,dsfont,physics}
\usepackage{tikz}
\usetikzlibrary{snakes}
\usepackage{accents}%
\providecommand{\U}[1]{\protect\rule{.1in}{.1in}}

\begin{document}
\title{Classical benchmarking for microwave quantum illumination}
\author{Athena Karsa}
\affiliation{Department of Computer Science, University of York, York YO10 5GH, UK}
\author{Stefano Pirandola}
\affiliation{Department of Computer Science, University of York, York YO10 5GH, UK}

\begin{abstract}
Quantum illumination (QI) theoretically promises up to a 6dB error-exponent advantage in target detection over the best classical protocol. The advantage is maximised by a regime which includes a very high background, which occurs naturally when one considers microwave operation. Such a regime has well-known practical limitations, though it is clear that, theoretically, knowledge of the associated classical benchmark in the microwave is lacking. The requirement of amplifiers for signal detection necessarily renders the optimal classical protocol here different to that which is traditionally used, and only applicable in the optical domain. This work outlines what is the true classical benchmark for microwave QI using coherent states, providing new bounds on the error probability and closed formulae for the receiver operating characteristic (ROC), for both optimal (based on quantum relative entropy) and homodyne detection schemes. An alternative source generation procedure based on coherent states is also proposed which demonstrates potential to reach classically optimal performances achievable in optical applications. The same bounds and measures for the performance of such a source are provided and its potential utility in the future of room temperature quantum detection schemes in the microwave is discussed.
\end{abstract}

\maketitle

\section{Introduction}\label{sec:intro}

Quantum illumination (QI)~\cite{lloyd2008enhanced,tan2008quantum,zhangexp,lopaevaexp,pirandola2018advances,shapiro2020quantum}  is a proposed protocol for quantum radar based on signal-idler entanglement which, theoretically, may achieve a 6 dB advantage in signal-to-noise ratio (SNR) (error-exponent) over its optimal classical counterpart, i.e., one without entanglement, operating at the same transmitted energy. This advantage persists even for weakly-reflecting targets embedded in a high background and despite the fact that the protocol itself is entanglement-breaking~\cite{zhangent}.

Original work on QI typically assumed operation at optical wavelengths where experimental tools are more readily available. However, at these wavelengths, one of the criteria for an optimal quantum advantage is not realistic: a high background. The natural solution was the theoretical extension of QI's operation to the microwave domain~\cite{barzanjeh2015microwave}, though practical difficulties here, including source-generation and signal-detection, are well-known~\cite{brandsemareadiness}. Despite this, recent initial microwave QI experiments~\cite{luong2020exp,barzanjeh2020microwave} have been carried out showing improved performances over their chosen classical comparison cases. This has been the subject of much debate, since these classical comparison cases are indeed different to the traditionally `optimal' one based on coherent states and their performances may indeed be viewed as sub-optimal. However, there are very few known methods for generating a low-energy semi-classical source for room temperature applications. Currently there are three potential procedures:
\begin{enumerate}
    \item Source is generated with an amplifier. A microwave coherent state at the single-photon level must first be generated at ultra-low temperature ($\sim 7$mK). Due to detector limitations and free-space loss, the signal must first be passed through an amplifier prior to probing a target region at $\sim 300$K. This process necessarily introduces noise to the state rendering the resultant source sub-optimal in the traditional sense.
    \item Source is generated without an amplifier. Recently, solid-state devices have been shown to be able to produce `microwave lasers', or masers, at room temperature. In QI applications, however, these sources must be heavily attenuated in order to achieve low enough photon numbers to form sensible comparisons with entanglement-based QI sources. In order to minimise noise and maintain an approximately coherent source, it is necessary to carry out this attenuation at cryogenic temperatures as will be seen in this work. Note that such a scheme has, as of yet, not been experimentally demonstrated but will be proposed as an alternative in this work with its efficacy studied.
    \item Source is generated without an amplifier or cryogenic attenuation. Such a protocol would require reliable low-energy microwave coherent state generation in addition to quantum-limited microwave detectors robust to thermal noise. This would ultimately yield the theoretically `optimal' classical source previously described, coinciding with what can be seen, as of yet, only in optical applications, however there is no currently known way to realise this.
\end{enumerate}
Note that the source generation method used in the prototypical experiment~\cite{barzanjeh2020microwave} was in fact a hybrid between procedures (1) and (2): a room temperature microwave source generated a weak coherent tone followed by a chain of low temperature attenuators which was then amplified to enable returning signal detection. Further, despite the fact that procedure (3) is impossible to carry out with current experimental capabilities, it persists to be assumed as the classical benchmark in almost all literature pertaining to microwave QI when benchmarking performances. While it is certainly valid and optimal in the optical regime, this does not translate to the microwave where it simply does not exist. Knowledge of the true, regime-dependent, classical benchmark is crucial in order to ascertain the existence of a quantum advantage.

Regardless of the classical benchmarking procedure considered, limitations on detectors pose problems for realistic implementation of coherent state illumination. Irrespective of how the source is generated and transmitted, use of a quantum detector is needed in order to receive such low-energy returning signals since homodyne detection does not work; a quantum detector design is required such that even if the input is coherent (classical), the radar system, as a whole, is still in fact quantum.

This paper outlines a true classical benchmark for microwave QI for room temperature applications, based on the fact that these techniques are, so far, the only known tools of generating an optimal classical source at the microwave. Sec.~\ref{sec:protocol} outlines two protocols for microwave QI using coherent states: the first, for a source generated with amplification; the second, proposed by this work, based on the output of a room temperature maser followed by heavy cryogenic attenuation. The tools of quantum hypothesis testing (QHT) are used in Secs.~\ref{sec:symmetric} and~\ref{sec:asymmetric} where formulae for the quantum Chernoff bound (QCB) and quantum relative entropy (QRE) are given, under symmetric and asymmetric considerations, respectively, yielding new error bounds for the microwave classical benchmark. In Sec.~\ref{sec:homodyne}, a protocol involving homodyne detection of the returning signal is considered with the resulting receiver operating characteristic (ROC) computed. In all cases the results for these new classical benchmarks are compared to the traditional one applicable only in the optical regime, constrained such that the total energy by which the target is irradiated is maintained. Up to here, this work's analyses are confined to regimes whereby the simultaneous study and comparison of classical benchmarks (1), (2) and (3) are possible. The sheer magnitude of the noise introduced by procedure (1) render the signal energy per mode so large that any quantum advantage would be diminished owing to the fact that the two-mode, signal-idler, entanglement correlations enabling the QI advantage becomes irrelevant at high brightness. Thus, in Sec.~\ref{sec:analysis} the results of Sec.~\ref{sec:classbench} are studied as the classical benchmark and compared to the performance of a two-mode squeezed vacuum (TMSV) source for entanglement-based QI, within the regime where such a protocol may be applied.

\section{Classical benchmark for microwave QI}\label{sec:classbench}

\subsection{Protocols for microwave QI using coherent states}\label{sec:protocol}

\subsubsection{Source generated with an amplifier}

For microwave QI experiments, the classical benchmark is obtained by replacing the QI source with one for coherent states in a fridge operating at $\simeq 7$mK. Amplifiers must be used to take the source out of this environment in order to probe and detect the presence or absence of a target existing at room temperature ($300$K), otherwise the SNR at detection will be too low. Such a process necessarily changes the returning state at the detector to one whose properties are typically very different to those used so far in classical benchmarking for QI. The protocol itself, illustrated in Fig.~\ref{fig:fridgeCS}, is outlined as follows:

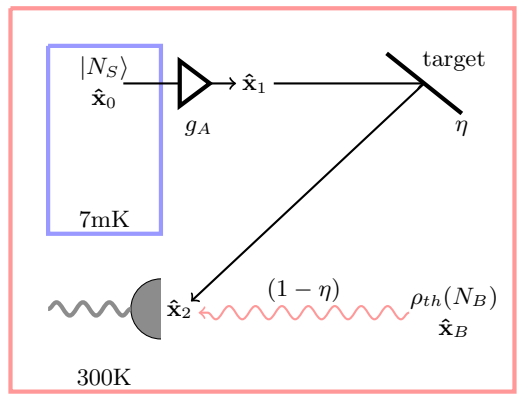
\begin{figure}[t!]
    \centering
\begin{tikzpicture}
\node at (0.75,-4.4) {$300$K};
\draw[-,red!40,ultra thick] (-0.5,0.5) -- (6.2,0.5) -- (6.2,-4.6) -- (-0.5,-4.6) -- (-0.5,0.5);
\node at (0.75,-2.3) {$7$mK};
\draw[-,blue!40,ultra thick] (0,0) -- (1.5,0) -- (1.5,-2.5) -- (0,-2.5) -- (0,0); 
\node at (0.75, -0.3) {$\ket{N_S}$};
\node at (0.75, -0.7) {$\mathbf{\hat{x}}_0$};
\draw[-,black,thick] (1,-0.5) -- (1.75,-0.5);
\draw[-,black,ultra thick] (1.75,-0.5) -- (1.75,-0.2) -- (2.15,-0.5) -- (1.75,-0.8) -- (1.75,-0.5);
\node at (2,-1.1) {$g_A$};
\draw[->, black, thick] (2.15,-0.5) --(2.5,-0.5);
\node at (2.75,-0.5) {$\mathbf{\hat{x}}_1$};
\draw[->, black, thick] (3,-0.5) --(5,-0.5);
\draw[-, black,ultra thick] (4.5,-0.1) --(5.5,-0.9);
\node at (5.4,-0.2) {target};
\node at (5.5,-1.1) {$\eta$};
\draw[->, black, thick] (5,-0.5) -- (1.9,-3.4);
\node at (1.75,-3.5) {$\mathbf{\hat{x}}_2$};
\draw[->,red!40, snake=snake, thick] (4.8,-3.55)-- (2,-3.55);
\node at (5.4, -3.35) {$\rho_{th}(N_B)$};
\node at (5.4, -3.75) {$\mathbf{\hat{x}}_B$};
\node at (3.4,-3.25) {$(1-\eta)$};
\begin{scope}
\clip (1.5,-4) rectangle (1,-3);
\draw[fill=black!45] (1.5,-3.5) circle(0.4);
\draw (6,2.75) -- (6,3.25);
\end{scope}
\draw[-,black!45, snake=snake, ultra thick] (1.1,-3.5)-- (0,-3.5);
\end{tikzpicture}
    \caption{Protocol for experimental microwave QI using a coherent state source generated with an amplifier. Each source mode is prepared in a coherent state $\ket{N_S}$ at 7mK which is passed through an amplifier of gain $g_A$ to probe a target with reflectivity $\eta$ residing at room temperature, 300K. The received signal is then mixed with the background which is in a thermal state $\rho_{th}(N_B)$.}
    \label{fig:fridgeCS}
\end{figure}

\begin{itemize}
    \item For the purpose of classical benchmarking, the input is prepared in a coherent state $\ket{N_S}^{\otimes M}$ with mean number of photons per mode, $M$, equal to $N_S$. Quadrature operators are given by $\mathbf{\hat{x}}_0 = (\hat{q_0},\hat{p_0})^{\mathrm{T}}$ with mean $ \mathbf{\bar{x}}_0 = (\sqrt{N_S},0)$ and covariance matrix $\mathbf{V}_0 = (1/2)\mathbf{1}_2$.
    \item Upon exiting the fridge to probe a target region at $T=300$K, the source must pass through an amplifier characterised by gain $g_{A} \geq 1$ which, assuming phase-preserving quantum-limited amplification,  transforms quadratures as
    \begin{equation}
        \mathbf{\hat{x}}_0 \rightarrow \mathbf{\hat{x}}_1 = \sqrt{g_{A}} \mathbf{\hat{x}}_0 + \sqrt{g_{A} - 1} \mathbf{\hat{x}}_{A},
    \end{equation}
    where $\mathbf{\hat{x}}_{A}$ are the quadrature operators associated with the amplifier. Rescaling the input as $\mathbf{\hat{x}}_0 \rightarrow \mathbf{\hat{x}}_0/\sqrt{g_{A}}$ yields as output
    \begin{equation}
        \mathbf{\hat{x}}_1 = \mathbf{\hat{x}}_0 + \sqrt{g_{A}-1} \mathbf{\hat{x}}_A
    \end{equation}
    with mean $\mathbf{\bar{x}}_1=\mathbf{\bar{x}}_0$ and covariance matrix $\mathbf{V}_1 = N_A\mathbf{1}_2$ where $N_A = N_B + \frac{1}{2}g_A$ is added number of photons added due to the amplifier, constituting classical noise. Note that $N_A \geq N_B$, where $N_A$ is the mean number of photons associated with the ambient background given by Planck's law, with equality when $g_A = 1$, a minimum. This state constitutes the source seen by the target; the target is irradiated by a displaced thermal state with higher total energy due to the combined photons from the original coherent state \emph{and} those added through the necessary use of an amplifier.
    \item The interaction of the source $\mathbf{\hat{x}}_1$ with the target may be modelled as a beamsplitter with transmissivity $\eta$. The returning signal at the receiver, $\mathbf{\hat{x}}_2$, is mixed with background photons constituting a thermal state $\rho_{th}(N_B)$ with $N_B/(1-\eta)$ average photons per mode and quadrature operators $\mathbf{\hat{x}}_B$ such that
    \begin{equation}
        \mathbf{\hat{x}}_2 = \sqrt{\eta}\mathbf{\hat{x}}_1 + \sqrt{1-\eta}\mathbf{\hat{x}}_B.
    \end{equation}
    This state has mean value $\mathbf{\bar{x}}_2=(\sqrt{\eta N_S},0)$ and variance,
    \begin{equation}
    \begin{split}
        \mathbf{V_2} &= \eta \left(\frac{1}{2} + N_A\right)\mathbf{1}_2 + (1-\eta) \left(\frac{1}{2} + \frac{N_B}{1-\eta}\right)\mathbf{1}_2\\
        &= \left(\frac{1}{2} + \eta N_A + N_B\right)\mathbf{1}_2.
        \end{split}
    \end{equation}
    \item Target detection is then reduced to discriminating between two hypotheses: $H_0$, target is absent and the received signal is just the thermal state $\mathbf{\hat{x}}_B$ with zero mean and covariance $\mathbf{V}_B = (N_B + 1/2)\mathbf{1}_2$; and $H_1$, target is present and the received signal is $\mathbf{\hat{x}}_2$.
\end{itemize}

\subsubsection{Source generated without an amplifier}

\begin{figure}[t!]
    \centering
\begin{tikzpicture}
\node at (0.75,-4.4) {$300$K};
\draw[-,red!40,ultra thick] (-0.5,0.5) -- (6.2,0.5) -- (6.2,-4.6) -- (-0.5,-4.6) -- (-0.5,0.5);
\node at (1.25,-2.3) {$T$ (K)};
\draw[-,blue!40,ultra thick] (0.5,0) -- (2,0) -- (2,-2.5) -- (0.5,-2.5) -- (0.5,0); 
\node at (0.05, -0.3) {$\ket{N_S}$};
\node at (0.05, -0.7) {$\mathbf{\hat{x}}_0$};
\draw[->, black, thick] (1.25,-1.5) --(1.25,-0.6);
\node at (1.25,-1.6) {$\bar{n}_{T}$};
\draw[-, black,ultra thick] (0.8,-0.1) --(1.7,-0.9);
\node at (1.7,-1.1) {$\phi$};
\draw[->, black, thick] (0.3,-0.5) --(2.5,-0.5);
\node at (2.75,-0.5) {$\mathbf{\hat{x}}_1$};
\draw[->, black, thick] (3,-0.5) --(5,-0.5);
\draw[-, black,ultra thick] (4.5,-0.1) --(5.5,-0.9);
\node at (5.4,-0.2) {target};
\node at (5.5,-1.1) {$\eta$};
\draw[->, black, thick] (5,-0.5) -- (1.9,-3.4);
\node at (1.75,-3.5) {$\mathbf{\hat{x}}_2$};
\draw[->,red!40, snake=snake, thick] (4.8,-3.55)-- (2,-3.55);
\node at (5.4, -3.35) {$\rho_{th}(N_B)$};
\node at (5.4, -3.75) {$\mathbf{\hat{x}}_B$};
\node at (3.4,-3.25) {$(1-\eta)$};
\begin{scope}
\clip (1.5,-4) rectangle (1,-3);
\draw[fill=black!45] (1.5,-3.5) circle(0.4);
\draw (6,2.75) -- (6,3.25);
\end{scope}
\draw[-,black!45, snake=snake, ultra thick] (1.1,-3.5)-- (0,-3.5);
\end{tikzpicture}
    \caption{Protocol for experimental microwave QI using a coherent state source generated without an amplifier. Each source mode is prepared in a coherent state $\ket{N_S}$ at 300K with $N_S \gg 1$, i.e., the output of a room temperature maser. Attenuation at temperature $T$ (K) with a beamsplitter with transmissivity $\phi$ mixes the source with environmental noise $\bar{n}_T$ yielding the final source which probes the target with reflectivity $\eta$ residing at room temperature, 300K. The received signal is then mixed with the background which is in a thermal state $\rho_{th}(N_B)$.}
    \label{fig:roomCS}
\end{figure}
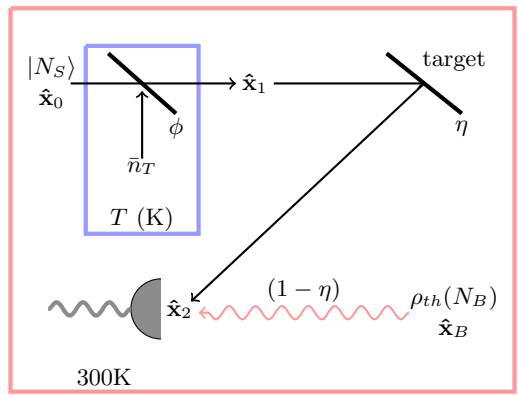

An alternative benchmark for microwave QI starts by generating a high energy microwave coherent state, such as the output of a room temperature maser. By passing this state through an ultra-cold beamsplitter, the source may be energetically diminished, providing a suitable benchmark for QI, while the necessary introduction of environmental noise through the beam splitting process is minimised by ensuring the local ambient temperature, and thus the local ambient background, is small. The protocol, illustrated in Fig.~\ref{fig:roomCS}, is outlined as follows:

\begin{itemize}
    \item As an alternative classical benchmark, the input is prepared in coherent state $\ket{N_S}^{\otimes M}$ with mean number of photons per mode $M$ equal to $N_S \gg 1$.  Quadrature operators are given by $\mathbf{\hat{x}}_0 = (\hat{q_0},\hat{p_0})^{\mathrm{T}}$ with mean $ \mathbf{\bar{x}}_0 = (\sqrt{N_S},0)$ and covariance matrix $\mathbf{V}_0 = (1/2)\mathbf{1}_2$.
    \item The source initially passes through a beamsplitter of transmissivity $\phi$ contained inside a fridge maintained at temperature $T$. Its specifications are such that $\phi \ll 1$ such that the resulting output state has a low energy suitable for use as a QI benchmark. The state transforms as $\mathbf{\hat{x}}_0 \rightarrow \mathbf{\hat{x}}_1$ with mean $\mathbf{\bar{x}}_1 = (\sqrt{\phi N_S}, 0)$ and covariance matrix $\mathbf{V_1} = (\bar{n}_T+1/2) \mathbf{1}_2$, where $\bar{n}_T = (\exp[\hbar \nu/k_B T -1] )^{-1}$.
    \item As previously described, the interaction of the source $\mathbf{\hat{x}}_1$ with the target may be modelled as a beamsplitter with transmissivity $\eta$ such that, at the receiver, we have the return state $\mathbf{\hat{x}}_2$ with mean $\mathbf{\bar{x}}_2 = (\sqrt{\eta \phi N_S}, 0)$ and covariance matrix $\mathbf{V_2} = (N_B + \eta \bar{n}_T+1/2) \mathbf{1}_2$ is the number of environmental photons per mode associated with the fridge operating at temperature $T$.
    \item As before, target detection is then reduced to discriminating between two hypotheses: $H_0$, target is absent and the received signal is just the thermal state $\mathbf{\hat{x}}_B$ with zero mean and covariance $\mathbf{V}_B = (N_B + 1/2)\mathbf{1}_2$; and $H_1$, target is present and the received signal is $\mathbf{\hat{x}}_2$.
    
\end{itemize}

\subsection{Classical benchmark for symmetric QHT}\label{sec:symmetric}

Under symmetric QHT one considers the minimisation of the average error probability in the discrimination of two quantum states. Applied to target detection, it provides a measure of the distinguishability of the returning states under each of the two alternative hypotheses: target present and target absent. The QCB provides an upper bound to the minimum error probability and, for Gaussian states, may be computed straightforwardly using closed formulae (see App. \ref{appQCB} for details).

\subsubsection{Source generated with an amplifier}

 Beginning by considering the first protocol for microwave coherent state generation with amplification, we can compute the QCB for such a source. Using an algebraic computation program, it can be found that the QCB for target detection using amplified microwave coherent states is given by
\begin{equation}
    P_{\mathrm{err}}^{\mathrm{CS, amp}} \leq \frac{1}{2 \xi_1}e^{ - M \eta N_S \xi_2 },
    \label{eq:QCBfridge}
\end{equation}
where
\begin{equation}
\begin{split}
    \xi_1 &= \Big( 1 + 2N_B (1 + N_B)  + \eta(N_A + 2 N_A N_B)\\
    &- 2\sqrt{N_B (1+N_B)(\eta N_A + N_B) (1 + \eta N_A + N_B)}  \Big)^{1/2},
    \end{split}
\end{equation}
and
\begin{equation}
     \xi_2 = \frac{ (\sqrt{N_B} - \sqrt{1+N_B}) (\sqrt{\eta N_A + N_B} - \sqrt{1 + \eta N_A + N_B}) }{   \sqrt{(1+N_B) (1 + \eta N_A + N_B)} -\sqrt{N_B (\eta N_A + N_B)} }.
\end{equation}

When amplifier noise $N_A \rightarrow 0$ the usual QI using coherent state bound is recovered which is valid in, for example, the optical regime, where amplifiers are not required,
\begin{equation}
    P_{\mathrm{err}}^{\mathrm{CS}} \leq \frac{1}{2}e^{ - M \eta N_S (\sqrt{N_B+1} - \sqrt{N_B})^2 }.
\end{equation}
This coincides with the performance of procedure (3) detailed in Sec.~\ref{sec:intro}. 

Further, in the limit where the background is very large, $N_B \gg 1$ we have that
\begin{equation}
    P_{\mathrm{err}}^{\mathrm{CS, amp}} = P_{\mathrm{err}}^{\mathrm{CS}} \approx  \frac{1}{2}e^{ - M \eta N_S/4 N_B},
\end{equation}
valid for any value of $N_A$, over which QI using a TMSV state has the well-established factor of 4 error-exponent advantage,
\begin{equation}
   P_{\mathrm{err}}^{\mathrm{TMSV}} \lesssim  \frac{1}{2}e^{ - M \eta N_S/N_B},
\end{equation}
in the limit of large background, $N_B \gg 1$, and small reflectivity, $\eta \ll 1$.

\begin{figure}[t]
    \centering
    \includegraphics[width=8.6cm]{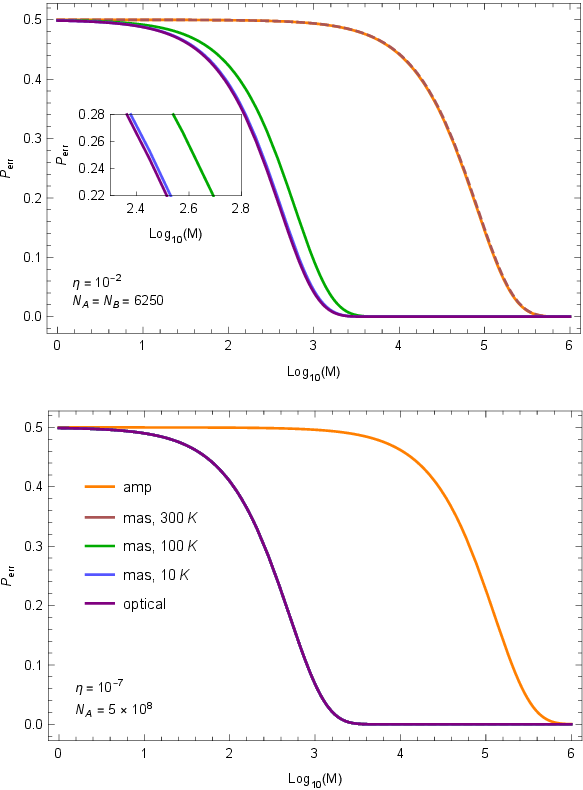}
    \caption{QCB for microwave QI classical benchmarks: 1) with the source generated inside a fridge followed by amplification (amp), and 2) with the source generated by a room temperature maser followed by attenuation at temperature $T$ (mas, $T$ K), compared to the optical coherent state performance. For the latter two cases we make the substitutions $\phi N_S \rightarrow N_S + N_A - \bar{n}_T$ and $N_S \rightarrow N_S + N_A$, respectively, with $N_S= 10^{-2}$. We assume the target is maintained at temperature $T=300$K, yielding $N_B=6250$. Upper panel: $N_A = N_B = 6250$, the minimum, with reflectivity $\eta=10^{-2}$ corresponding to a target range of $0.25$m assuming receiver collecting area of $0.1$m$^2$. Lower panel: $N_A = 5 \times 10^8$, a typical experimental value, with reflectivity $\eta=10^{-7}$ corresponding to a target range $\simeq 80$m.  Note with lower $N_A$ and thus signal energy, (mas, $300$ K) overlaps with amp (dashed line) highlighting value of cryogenic attenuation. At higher $N_A$ all of the (mas, $T$ K) plots here overlap with that for the optical coherent state due to the magnitude of $N_A$ and resultant energies.}
    \label{fig:QCBplot}
\end{figure}

\subsubsection{Source generated without an amplifier}

The second, alternative, protocol for the generation of low-energy microwave coherent sources for the purposes of QI benchmarking has not been experimentally demonstrated so far. It requires precise use of a room temperature maser alongside controlled beam splitting at cryogenic temperatures in order to create a suitable state for illumination. As done previously, using an algebraic computation program, it can be found that the QCB for target detection using microwave coherent states generated in this manner is given by

\begin{equation}
    P_{\mathrm{err}}^{\mathrm{CS, mas}} \leq \frac{1}{2 \chi_1}e^{ - M \eta N_S \phi \chi_2 },
        \label{eq:QCBroom}
\end{equation}
where
\begin{equation}
\begin{split}
    \chi_1 &= \Big( 1 + 2N_B (1 + N_B)  + \eta(\bar{n}_T + 2 \bar{n}_T N_B)\\
    &- 2\sqrt{N_B (1+N_B)(\eta \bar{n}_T + N_B) (1 + \eta \bar{n}_T + N_B)}  \Big)^{1/2},
    \end{split}
\end{equation}
and
\begin{equation}
     \chi_2 = \frac{ (\sqrt{N_B} - \sqrt{1+N_B}) (\sqrt{\eta \bar{n}_T + N_B} - \sqrt{1 + \eta \bar{n}_T + N_B}) }{   \sqrt{(1+N_B) (1 + \eta \bar{n}_T + N_B)} -\sqrt{N_B (\eta \bar{n}_T + N_B)} }.
\end{equation}
Notice that the QCB for a room temperature generated source, Eq.~(\ref{eq:QCBroom}), is a very similar form to that of one generated with amplification, Eq.~(\ref{eq:QCBfridge}). The change of parameters $\xi_{1(2)} \rightarrow \chi_{1(2)}$ is done by replacing the added noise due to amplification $N_A \rightarrow \bar{n}_T$, the number of photons per mode associated to the fridge operating at temperature $T$ K. Further, there is an additional factor $\phi$, the transmissivity of the cryogenic beamsplitter inside the fridge used to create the low-energy QI source in the error-exponent, essentially modulating the SNR by that same amount.

As the fridge temperature $T\rightarrow 0$ K, the added noise $\bar{n}_T\rightarrow 0$ as well. Imposing this limit along with that for $N_B \gg 1$, the error probability becomes
\begin{equation}
    P_{\mathrm{err}}^{\mathrm{CS, mas}} \simeq \frac{1}{2}e^{ - M \eta \phi N_S /4 N_B },
\end{equation}
and such a source generated at room temperature performs as the well-known classical benchmark for the optical regime. However, in this scenario, such a performance may be achieved dependent on the temperature of attenuation.

Fig.~\ref{fig:QCBplot} shows plots the total error probability, using the QCB, for the microwave QI classical benchmarks: (1) with the source generated inside a fridge followed by amplification (amp), and (2) with the source generated by a room temperature maser followed by attenuation at temperature $T$ (mas, $T$ K), compared to the un-amplified optical coherent state performance which would coincide with protocol (3) from Sec.~\ref{sec:intro} if it were possible (at the microwave). In order to maintain the overall energy by which the target is irradiated, the substitutions $\phi N_S \rightarrow N_S + N_A - \bar{n}_T$ and $N_S \rightarrow N_S + N_A$ is made for the latter two, un-amplified, cases. When the source is generated by a maser followed by cryogenic attenuation, the performance closely coincides with that of the coherent state operating in the optical domain at only $10$K.

\subsection{Classical benchmark for asymmetric QHT}\label{sec:asymmetric}

Asymmetric QHT, rather than minimising the total average error probability, allows for some small, fixed type-I (false alarm) error, $P_{\mathrm{fa}} < \epsilon$, in an attempt to further minimise the type-II (missed detection) error, $P_{\mathrm{md}}$. Following quantum Stein's lemma, the quantum relative entropy (QRE) and the quantum relative entropy variance (QREV) give the optimal decay rate of the type-II error in this scenario (see App. \ref{appQRE} for details). An alternative approach using the quantum Hoeffding bound~\cite{audenaert2008asymptotic,spedalieri2014asymmetric} is not considered here.

\begin{figure}[t]
    \centering
    \includegraphics[width=8.6cm]{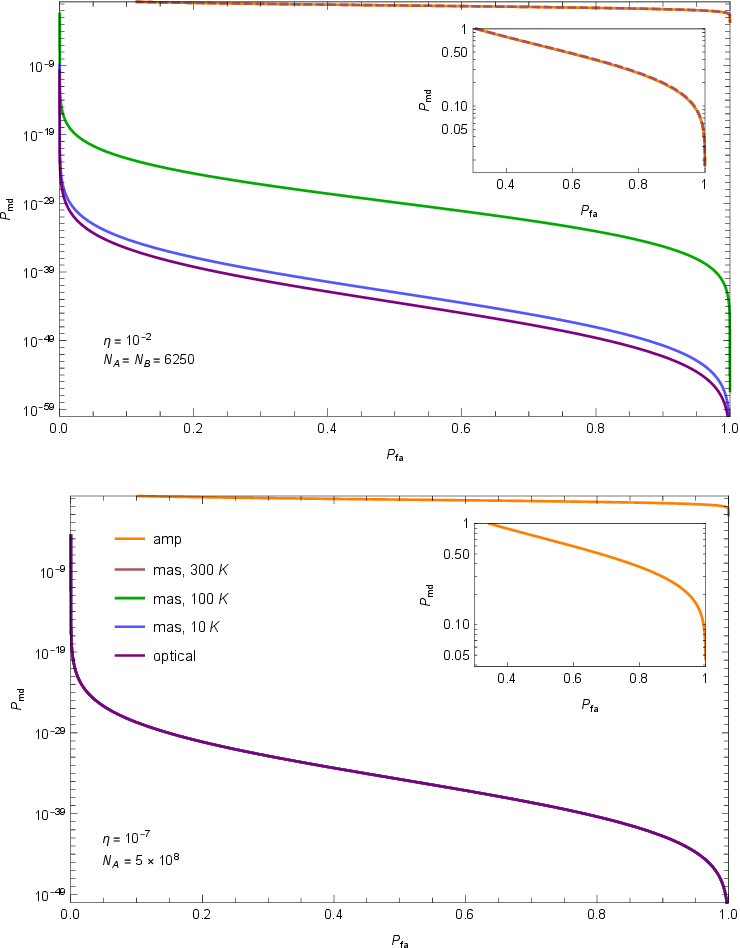}
    \caption{ROC curves generated using the QRE and QREV for microwave QI classical benchmarks: 1) with the source generated inside a fridge followed by amplification (amp), and 2) with the source generated by a room temperature maser followed by attenuation at temperature $T$ (mas, $T$ K), compared to the optical coherent state performance. For the latter two cases we make the substitutions $\phi N_S \rightarrow N_S + N_A - \bar{n}_T$ and $N_S \rightarrow N_S + N_A$, respectively, with $N_S= 10^{-2}$. We assume the target is maintained at temperature $T=300$K, yielding $N_B=6250$. Upper panel: $N_A = N_B = 6250$, the minimum, with reflectivity $\eta=10^{-2}$ corresponding to a target range of $0.25$m assuming receiver collecting area of $0.1$m$^2$. Lower panel: $N_A = 5 \times 10^8$, a typical experimental value, with reflectivity $\eta=10^{-7}$ corresponding to a target range $\simeq 80$m. In all cases we set $M=10^5$. Note with lower $N_A$ and thus signal energy, (mas, $300$ K) overlaps with amp (dashed line) highlighting value of cryogenic attenuation. At higher $N_A$ all of the (mas, $T$ K) plots here overlap with that for the optical coherent state due to the magnitude of $N_A$ and resultant energies.}
    \label{fig:QREROCplot}
\end{figure}

\subsubsection{Source generated with amplification}

It can be found that the QRE and QREV for the target detection using microwave coherent states generated with amplification are given by
\begin{equation}
\begin{split}
    D^{\mathrm{CS,amp}} &= \frac{1}{2}\Bigg( (1 + 2N_B + 2\eta N_S) \ln \left(1+\frac{1}{\eta N_A + N_B} \right)\\
    &- (1+2 N_B) \ln \left(1 + \frac{1}{N_B} \right) \\
    &+ \ln \left( \frac{(\eta N_A + N_B) (1 + \eta N_A + N_B )}{N_B(1+N_B)}\right)\Bigg),
    \end{split}
\end{equation}
and
\begin{equation}
\begin{split}
    V&^{\mathrm{CS,amp}} = N_B (1+N_B)\ln ^2 \left(1 + \frac{1}{N_B} \right)\\
    - 2 N_B&(1+N_B)\ln \left(1+ \frac{1}{N_B} \right) \ln \left(1 + \frac{1}{\eta N_A + N_B} \right)\\
    + (N_B&(1+N_B) + \eta N_S + 2 \eta N_S N_B) \ln^2 \left( 1 + \frac{1}{\eta N_A +N_B} \right),
    \end{split}
\end{equation}
respectively.
As in the symmetric case, when amplifier noise $N_A \rightarrow 0$, these expressions recover the known quantities for a coherent state transmitter~\cite{wilde2017qht,karsa2020gensource} given by
\begin{equation}
    D^{\mathrm{CS}} = \eta N_S \ln \left(1 + \frac{1}{N_B} \right),
\end{equation}
and
\begin{equation}
    V^{\mathrm{CS}} = \eta N_S ( 2N_B + 1) \ln^2 \left( 1 + \frac{1}{N_B}\right).
\end{equation}

\subsubsection{Source generated without amplification}

Alternatively, for the coherent state source generated without amplification, the QRE and QREV for target detection are given by
\begin{equation}
\begin{split}
    D^{\mathrm{CS,mas}} &= \frac{1}{2}\Bigg( (1 + 2N_B + 2\eta \phi N_S) \ln \left(1+\frac{1}{\eta \bar{n}_T + N_B} \right)\\
    &- (1+2 N_B) \ln \left(1 + \frac{1}{N_B} \right) \\
    &+ \ln \left( \frac{(\eta \bar{n}_T + N_B) (1 + \eta \bar{n}_T + N_B )}{N_B(1+N_B)}\right)\Bigg),
    \end{split}
\end{equation}
and
\begin{equation}
\begin{split}
    V&^{\mathrm{CS,mas}} = N_B (1+N_B)\ln ^2 \left(1 + \frac{1}{N_B} \right)\\
    - 2 N_B&(1+N_B)\ln \left(1+ \frac{1}{N_B} \right) \ln \left(1 + \frac{1}{\eta \bar{n}_T + N_B} \right)\\
    + (N_B&(1+N_B) + \eta \phi N_S + 2  \eta \phi N_S N_B) \ln^2 \left( 1 + \frac{1}{\eta \bar{n}_T +N_B} \right),
    \end{split}
\end{equation}
respectively.
As in the case for the QCB, the forms of QRE and QREV for the two coherent state sources are very similar with the replacement $N_A \rightarrow \bar{n}_T$ and an additional factor $N_S \rightarrow \phi N_S$ due to the action of a beamsplitter.

Together, combined with the constraint that $P_{\mathrm{fa}}\leq \epsilon$, these can be used to compute the corresponding probability of missed detection by
\begin{equation}
P_{\text{md}}=\exp\Big\{  -\Big[  MD +\sqrt{MV }\Phi^{-1}(\epsilon) + \mathcal{O}(\log M)\Big]  \Big\}
,\label{typeIIerrorexpform}%
\end{equation}
enabling us to calculate the relevant receiver operating characteristic (ROC) curves.

Fig.~\ref{fig:QREROCplot} shows plots the ROCs, based on the QRE and QREV, for microwave QI classical benchmarks: (1) with the source generated inside a fridge followed by amplification (amp), and (2) with the source generated by a room temperature maser followed by attenuation at $T$K (mas, $T$K), compared to the optical coherent state performance which would coincide with protocol (3) from Sec.~\ref{sec:intro} if it were possible (at the microwave). For the latter two cases we make the substitutions $\phi N_S \rightarrow N_S + N_A - \bar{n}_T$ and $N_S \rightarrow N_S + N_A$, respectively, to ensure that the total energy by which the target is irradiated is maintained. As seen in the symmetric case with the QCB, attenuating the maser source at a cryogenic temperature achieves a ROC closely coinciding with that of the optical coherent state.

\subsection{ROC with homodyne detection}\label{sec:homodyne}

In the case of coherent states with homodyne detection (combined with coherent integration and binary decision-making on the measurement results) the ROC is given by combining the following expressions
\begin{align}
P_{\text{fa}}^{\text{hom}}(x) &  =\frac{1}{2}\operatorname{erfc}%
\left[  \frac{x}{\sqrt{2 M\lambda_0}}\right] ,\label{eqq1}\\
P_{\text{md}}^{\text{hom}}(x) &  =\frac{1}{2}\operatorname{erfc}\left[ \frac{M \sqrt{\eta N_{S}} - x}{\sqrt{2 M \lambda_1}}\right]    ,\label{eqq2}%
\end{align}
where $\operatorname{erfc}(z):=1-2\pi^{-1/2}\int_{0}^{z}\exp(-t^{2})dt$ is the
complementary error function~\cite{karsa2020gensource}. 

\begin{figure}[hbt!]
    \centering
    \includegraphics[width=8.6cm]{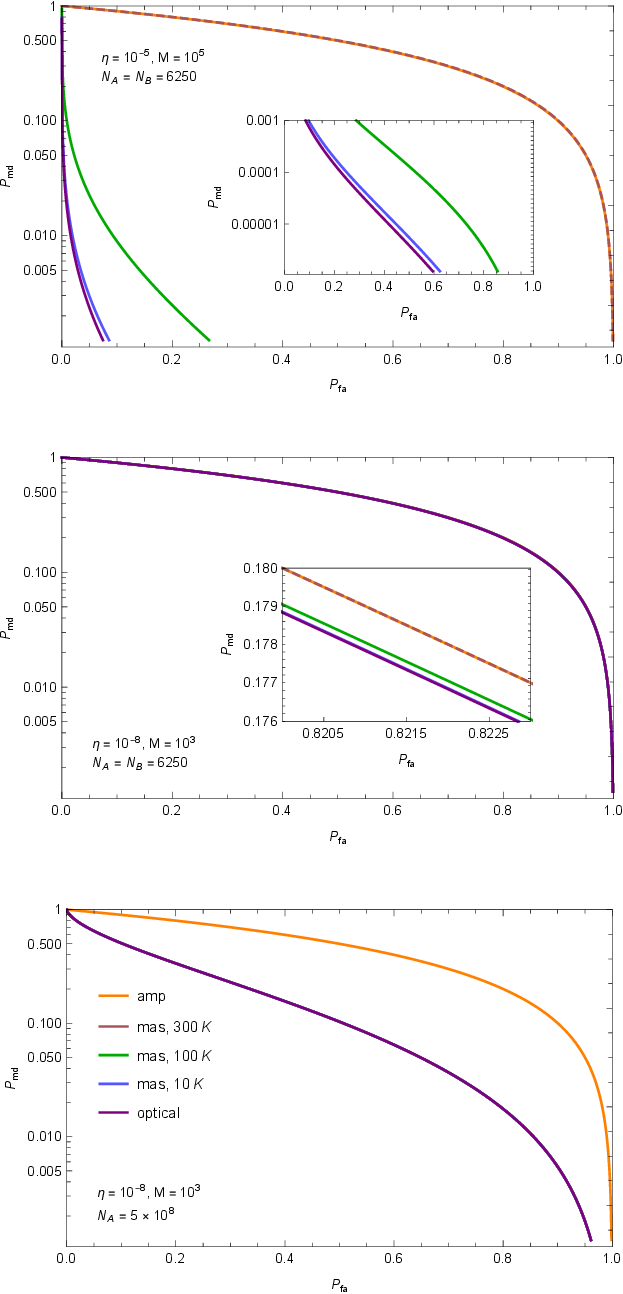}
    \caption{ROC curves for coherent state microwave illumination with homodyne detection: 1) with the source generated inside a fridge followed by amplification (amp), and 2) with the source generated by a room temperature maser followed by attenuation at temperature $T$ (mas, $T$ K), compared to the optical coherent state. For the latter two cases we make the substitutions $\phi N_S \rightarrow N_S + N_A - \bar{n}_T$ and $N_S \rightarrow N_S + N_A$, respectively, with $N_S= 10^{-2}$. Assume target is maintained at temperature $T=300$K, yielding $N_B=6250$. Upper panel: $N_A = N_B = 6250$, the minimum, with $\eta=10^{-5}$ corresponding to a target range $\simeq 8$m assuming receiver collecting area of $0.1$m$^2$ and $M=10^5$. Middle panel: $N_A = N_B = 6250$, the minimum, with $\eta=10^{-8}$ and $M=10^3$. For these two panels, with lower $N_A$ and thus signal energy, (mas, $300$ K) overlaps with amp (dashed line) highlighting value of cryogenic attenuation. Lower panel: $N_A = 5 \times 10^8$, a typical experimental value, with $\eta=10^{-8}$ and $M=10^3$. Note that all of the (mas, $T$ K) plots here overlap with that for the optical coherent state due to the magnitude of $N_A$ and resultant energies. Middle and lower panels correspond to a target range $\simeq 250$m assuming receiver collecting area of $0.1$m$^2$.}
    \label{fig:HomodyneROCplot}
\end{figure}

Regardless of means of source generation, and in both optical and microwave applications, the mean value of the returning signal, in the case where the target is present, is the same and equal to $(\sqrt{\eta N_S,0})$. Further, in the case of a null hypothesis, for Eq.~(\ref{eqq1}) we have that $\lambda_0 = N_B+1/2$ which holds in all considered classical benchmarking protocols. However, the effect of both the amplification and attenuation in the two protocols considered is to introduce additional noise to the system prior to target illumination. Thus for Eq.~(\ref{eqq2}) we have that $\lambda_1 \rightarrow \lambda_{\mathrm{amp}} = \eta N_A + N_B + 1/2$ in the amplified case, $\lambda_1 \rightarrow \lambda_{\mathrm{mas}} = \eta \bar{n}_T + N_B + 1/2$ in the case of the cryogenically attenuated maser, and $\lambda_1 \rightarrow \lambda_{\mathrm{opt}} = N_B + 1/2$ in the (optimal) optical case. Then, using the appropriate values for variances, Eq.~(\ref{eqq1}) can be inverted and substituted into Eq.~(\ref{eqq2}) to derive the corresponding ROC in each of the considered protocols.

Fig.~\ref{fig:HomodyneROCplot} plots the ROC curves for coherent state microwave illumination with homodyne detection: (1) with the source generated inside a fridge followed by amplification (amp), and (2) with the source generated by a room temperature maser followed by attenuation at temperature $T$ (mas, $T$ K), compared to the optical coherent state performance would coincide with protocol (3) from Sec.~\ref{sec:intro} if it were possible (at the microwave). For the latter two cases we make the substitutions $\phi N_S \rightarrow N_S + N_A - \bar{n}_T$ and $N_S \rightarrow N_S + N_A$, respectively, to maintain the total energy incident on the target. As seen with previous results, the proposed technique for source generation based on the output of a room temperature maser performs very closely to the optimal, optical coherent state as long as the attenuating temperature is low.

\section{Using the new classical benchmark}\label{sec:analysis}

The purpose of the comparisons, seen in Figs.~\ref{fig:QCBplot},~\ref{fig:QREROCplot} and~\ref{fig:HomodyneROCplot} is to allow for the proper comparison of the three classical benchmarks first outlined in Sec.~\ref{sec:intro} by constraining the total energy by which the target is irradiated. Procedure (1) is responsible for setting this constraint due to the magnitude of the noise introduced by the room temperature amplifier. At a minimum, this is equal to the ambient background which, for room temperature applications ($T=300$ K), is given by $N_A = N_B \simeq 6250$. This further determines the necessary values of other parameters such as reflectivity $\eta$ which corresponds to target range, to ultimately allow for the performance comparison.

Of course, such high signal energies per mode are not a valid regime for entanglement-based QI since the correlations which quantify the amount of entanglement are maximised for $N_S\ll 1$; they become irrelevant for $N_S\gg 1$\cite{karsa2020gensource}. Limiting our attentions to only procedures (2), room temperature maser source followed by cryogenic attenuation, and (3), currently only possible in the optical domain, a final comparison can be made to the entanglement-based QI using a TMSV source.

\begin{figure}[hbt!]
    \centering
    \includegraphics[width=\linewidth]{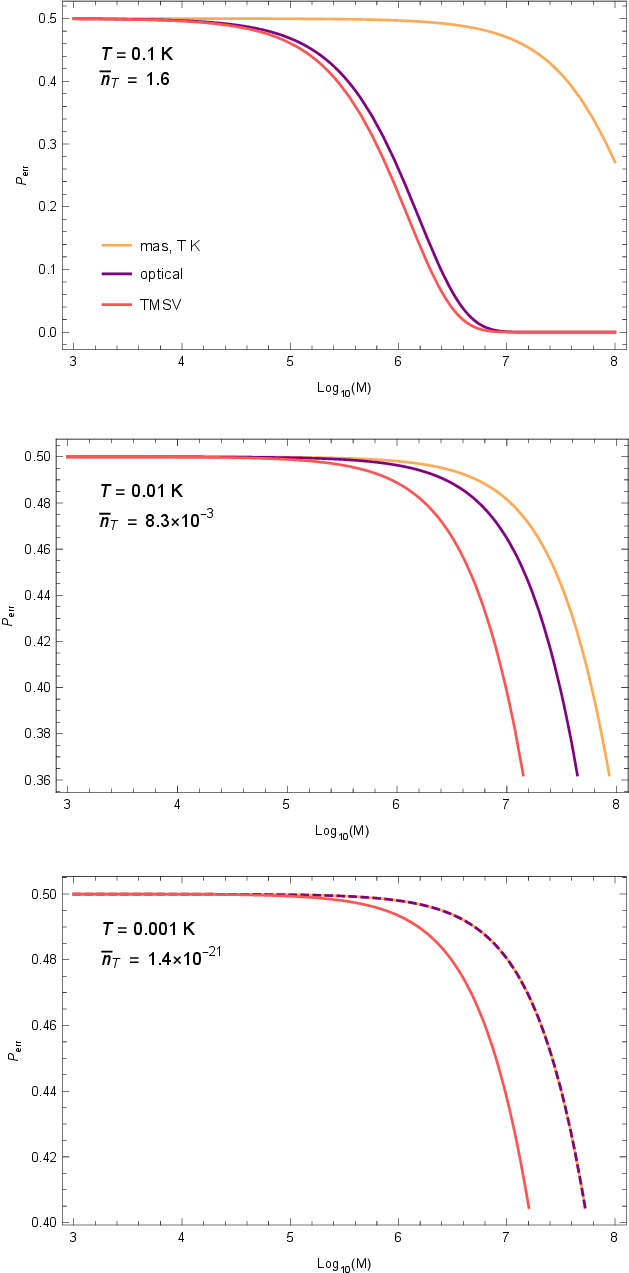}
    \caption{QCB for TMSV microwave QI compared with the new classical benchmarks. Considered are 1) the source generated by a room temperature maser followed by attenuation at temperature $T$ (mas, $T$ K), 2) comparison with to the optical coherent state performance, and 3) the TMSV. For the latter two cases we make the substitutions $N_S \rightarrow N_S + \bar{n}_T$ with $N_S= 10^{-2}$. We assume the target is maintained at temperature $T=300$K, yielding $N_B=6250$. Upper panel: $T=0.1$ K $\Rightarrow \bar{n}_T=1.6$. Middle panel: $T=0.01$ K $\Rightarrow \bar{n}_T=8.3 \times 10^{-3}$. Lower panel: $T=0.001$ K $\Rightarrow \bar{n}_T=1.4 \times 10^{-21}$. We set $\eta=10^{-2}$ corresponding to a target range of $0.25$m assuming receiver collecting area of $0.1$m$^2$. As expected, procedure (2) for practical microwave QI benchmarking coincides with the theoretically optimal, optical, benchmark (equivalent to procedure (3)) at very low temperatures.}
    \label{fig:classbenchcompTMSV}
\end{figure}

Fig.~\ref{fig:classbenchcompTMSV} plots the QCBs as a function of total number of probings $M$ for the classical benchmark (2) (mas, $T$ K)\clearpage 

\noindent the optical coherent state (3) and the TMSV QI protocol (see Ref.~\cite{karsa2020gensource} for full details). Energetic constraints are determined by (mas, $T$ K) whereby the amplification process alters the total energy by which the target is irradiated as $N_S\rightarrow N_S + \bar{n}_T$, and these are the substitutions made for energy in the latter two cases. Note that the values plotted for the TMSV source are exact and valid for all parameter values, i.e., no assumptions have been made as to their relative magnitudes.

\begin{figure}[t!]
    \centering
    \includegraphics[width=\linewidth]{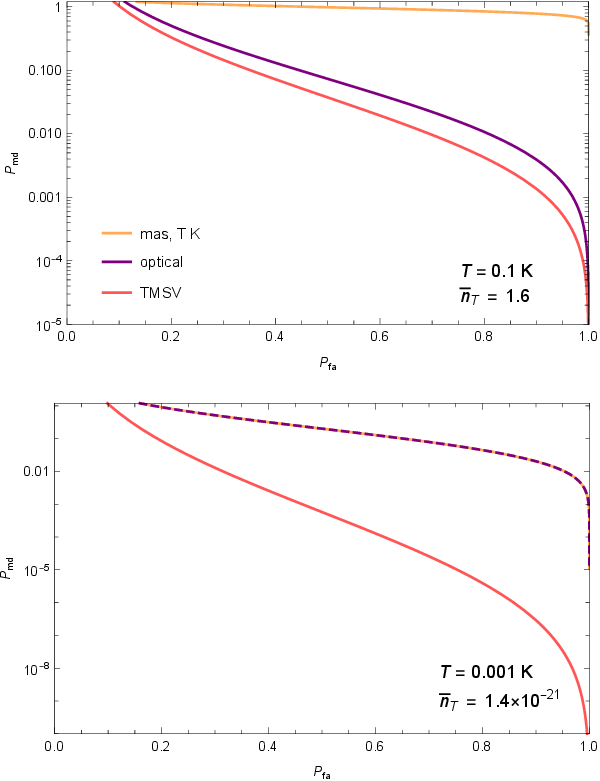}
    \caption{ROCs for TMSV microwave QI compared with the new classical benchmarks. Considered are 1) the source generated by a room temperature maser followed by attenuation at temperature $T$ (mas, $T$ K), 2) comparison with to the optical coherent state performance, and 3) the TMSV. For the latter two cases we make the substitutions $N_S \rightarrow N_S + \bar{n}_T$ with $N_S= 10^{-2}$. We assume the target is maintained at temperature $T=300$K, yielding $N_B=6250$. Upper panel: $T=0.1$ K $\Rightarrow \bar{n}_T=1.6$. Lower panel: $T=0.001$ K $\Rightarrow \bar{n}_T=1.4 \times 10^{-21}$. We set $\eta=10^{-2}$ corresponding to a target range of $0.25$m assuming receiver collecting area of $0.1$m$^2$. As expected, procedure (2) for practical microwave QI benchmarking coincides with the theoretically optimal, optical, benchmark (equivalent to procedure (3)) at very low temperatures.}
    \label{fig:classbenchcompTMSVrelent}
\end{figure}

Fig.~\ref{fig:classbenchcompTMSVrelent} shows the ROCs, based on the QRE and QREV (see Ref.~\cite{karsa2020gensource} for full details), for the same sources and under the same energetic constraints as Fig.~\ref{fig:classbenchcompTMSV}. As for the TMSV QCB, here we use the exact expressions for QRE and QREV in the computation of the ROC such that it is valid for any choice of parameter values.

 Results show, as expected, that provided attenuation occurs at small enough $T$, added noise may be diminished such that the practical coherent state source generation (2) may achieve performances coinciding with the `optimal' classical benchmark (3), currently only achievable at the microwave. Further, in the regimes considered the TMSV state retains its quantum advantage.

\section{Concluding remarks}

The aim of this work is to outline a true classical benchmark for microwave QI. Current experimental abilities (also limitations) mean that the only way to generate an `optimal' classical state in the microwave regime is also one which ultimately renders the source `sub-optimal' compared to traditional notions of the term, though these are based on optical applications where the impeding issues do not exist. 

As of yet, the only readily available technique for generating microwave coherent states for room temperature applications (for the purposes of QI benchmarking) requires the use of amplifiers to enable detection of a transmitted signal from low-temperature source environment to the target region. This paper outlines the practical protocol for microwave QI using coherent states, and computes the appropriate performance metrics, which may be used for comparisons, in terms of the total noise added due to the necessary use of amplifiers in the protocol. A new bound on the error probability for classical benchmarking in the microwave is given, alongside closed formulae for the ROCs using both the QRE for optimal performances and with homodyne detection, showing the inherent sub-optimality of such a procedure.

A further method is proposed based on a source generated via cryogenically attenuating the output of a room temperature maser. Then, by choosing appropriate levels of attenuation within a cold enough environment one could potentially generate any appropriately specified source for the given detection problem, taking into account detector limitations, forgoing the need for amplification. \clearpage

Bounds on the target detection error probability using the QCB are given alongside ROCs based on the QRE and a scheme based on homodyne detection. Results show that such a protocol shows promise in being able to act as an `optimal' one in the microwave, demonstrating a performance coinciding with optical coherent states provided the attenuation occurs at a low enough temperature to minimise added noise. 

Recent developments in continuous wave room temperature masers using optically pumped nitrogen-vacancy (NV$^{-}$) defect centres in diamond~\cite{breeze2018continuous} and a pentacene-doped crystal~\cite{wu2020room} give promise to the eventual realisation of such a procedure. Operating at $1.45$ GHz and reporting a peak output power of $\sim -90$dBm and $-25$dBm, respectively, yielding a rate of $\sim 10^{12}$ and $\sim 10^{18}$ photons per second, such devices could potentially be used for microwave coherent state source generation provided their noise temperatures are kept low.

In Sec.~\ref{sec:analysis} the results from preceding sections were employed as the classical benchmark to compare with TMSV performances, verifying results hold even in regimes where a quantum advantage exists. Explicitly, this regime is one of low brightness ($N_S \ll 1$), high background ($N_B \gg 1$), naturally satisfied in the microwave domain, and low reflectivity ($\eta \ll 1$). There is, however, a trade-off between the first and last requirements - particularly with respect to the fact that $\eta$ determines to what problems QI may be applied as it incorporates the target range which is, by far, the most dominant loss factor. Confined to a single-use protocol, there is no theoretical limitation to the coherent state signal energy such that propagation losses may be overcome to still yield a result. For a protocol based on entanglement, having to keep $N_S$ small means that the burden of overcoming such losses is shifted to the number of uses $M$, rendering the value so large that, at least in the near-term, may be experimentally unfeasible.

Although this work's focus has been on classical benchmarking for microwave QI, the proposed alternative method which may prove optimal, as in the optical regime, relies on technologies which are fundamentally quantum. The underlying process in these room temperature maser devices is the optical pumping of quantum spin states that are engineered to exist within the relevant material. Furthermore, for this to be effective and have signal states with quantum-limited noise at the microwave, the use of cryogenic temperature appears to be essential. Yet, any illumination device comprises two distinct components: the source generator and the signal detector. The enhancement of either or both of these through quantum means would ultimately yield a quantum-enhanced device. Thus one could argue that the classical benchmark for microwave QI proposed here could play two roles: first, at short ranges depending on experimental capabilities, as an optimal classical benchmark for evaluating a quantum advantage; and second, at medium-long ranges, a potential QI device which simply does not use entanglement.

\textbf{Acknowledgments.}~This work has been funded by the European Union's Horizon
2020 Research and Innovation Action under grant
agreement No. 862644 (FET-Open project: Quantum readout techniques
and technologies, QUARTET). A.K. acknowledges sponsorship by EPSRC Award No. 1949572 and Leonardo UK. The authors would like to thank J. Fink, S. Paraoanu and K. Petrovnin for discussions and feedback.

\appendix
\section{Gaussian state hypothesis testing and detection}

Consider the task of discriminating between two arbitrary $N$-mode Gaussian states, $\hat{\rho}_{0}%
(\mathbf{x}_{0},\mathbf{V}_{0})$ and $\hat{\rho}_{1}(\mathbf{x}_{1}%
,\mathbf{V}_{1})$, with mean $\mathbf{x}_{i}$ and CM $\mathbf{V}_{i}$ with
quadratures $\mathbf{\hat{x}}=\left(  \hat{q}_{1},\hat{p}_{1},\dots,\hat
{q}_{N},\hat{p}_{N}\right)  ^{T}$ and associated symplectic form
\begin{equation}
\mathbf{\Omega}=\bigoplus_{k=1}^{N}%
\begin{pmatrix}
0 & 1\\
-1 & 0
\end{pmatrix}
.
\end{equation}
Consider $M$ identical copies $\hat{\rho}%
_{i}^{\otimes M}$ of the state $\hat{\rho}_{i}$ encoding the classical
information bit $i\in\{0,1\}$. The optimal measurement for the discrimination
is the dichotomic positive-operator valued measure
(POVM)~\cite{helstrom1969quantum} $E_{0}=\Pi(\gamma_{+})$, $E_{1}=1-\Pi
(\gamma_{+})$, where $\Pi(\gamma_{+})$ is the projector on the positive part
$\gamma_{+}$ of the non-positive Helstrom matrix $\gamma:=\hat{\rho}_{0}^{\otimes M}-\hat{\rho}_{1}^{\otimes M}$. This allows for $\hat{\rho}_{0}$ and $\hat{\rho}_{1}$ to be
discriminated with a \emph{minimum} error probability given by the Helstrom
bound, $P_{\text{err}}^{\mathrm{min}}=\left[  1-D(\hat{\rho}_{0}^{\otimes M},\hat{\rho
}_{1}^{\otimes M})\right]  /2$, where $D$ is the trace distance~\cite{watrous2018quantum}.

This forms the symmetric approach to QHT where one's aim is to obtain a global minimization over all errors, irrespective of
their origin. In this case, one considers the minimization of the average
error probability
\begin{equation}
P_{\text{err}}:=P(H_{0})P(H_{1}|H_{0})+P(H_{1})P(H_{0}|H_{1}),\label{avgP}%
\end{equation}
where $P(H_{0})$ and $P(H_{1})$ are the prior probabilities associated with
the two hypotheses.

In asymmetric QHT, we wish to minimize one type of error as much as possible
while allowing for some flexibility on the other. Consider again $M$ identical
copies of the state $\hat{\rho}_{i}$ ($\hat{\rho}_{i}^{\otimes M}$), encoding
the classical bit $i\in\{0,1\}$. As in the symmetric case, the optimal choice
of measurement is a dichotomic POVM $\{E_{0},E_{1}\}$. From the binary
outcome, we can define the two types of error, i.e., the type-I (false alarm)
error
\begin{equation}
P_{\text{fa}}:=P(H_{1}|H_{0})=\Tr\left(  E_{1}\hat{\rho}_{0}^{\otimes
M}\right)  ,
\end{equation}
and the type-II (missed detection) error
\begin{equation}
P_{\text{md}}:=P(H_{0}|H_{1})=\Tr\left(  E_{0}\hat{\rho}_{1}^{\otimes
M}\right)  .
\end{equation}
These probabilities are dependent on the number $M$ of copies and, for $M\gg1$,
they both tend to zero, i.e.,
\begin{equation}
P_{\text{fa}}\simeq e^{-\alpha_{R}M},~P_{\text{md}}\simeq e^{-\beta_{R}M},
\end{equation}
where we define the `error-exponents' or `rate limits' as
\begin{equation}
\alpha_{R}=-\lim_{M\rightarrow+\infty}\frac{1}{M}\ln P_{\text{fa}},
\end{equation}
\vspace{-0.5cm}
\begin{equation}
\beta_{R}=-\lim_{M\rightarrow+\infty}\frac{1}{M}\ln P_{\text{md}}.
\end{equation}

\subsection{Quantum Chernoff bound for symmetric QHT}\label{appQCB}


Because this is difficult to compute analytically, the Helstrom bound is often
replaced with approximations such as the quantum Chernoff bound (QCB)~\cite{QCB},
\begin{equation}
P_{\text{err}}^{\mathrm{min}}\leq P_{\text{err}}^{\mathrm{QCB}}:=\frac{1}%
{2}\left(  \inf_{0\leq s\leq1}C_{s}\right)^{M}  ,~~C_{s}:=\Tr\left(  \hat{\rho
}_{0}^{s}\hat{\rho}_{1}^{1-s}\right).\label{QCBtext}
\end{equation}
Minimization of the $s$-overlap $C_{s}$ occurs over all $0\leq s\leq1$.
Forgoing minimization and setting $s=1/2$ one defines a simpler, though
weaker, upper bound, also known as the quantum Bhattacharyya bound (QBB)~\cite{RMP}%
\begin{equation}
P_{\text{err}}^{\mathrm{QBB}}:=\frac{1}{2}\Tr\left(  \sqrt{\hat{\rho}_{0}%
}\sqrt{\hat{\rho}_{1}}\right)^{M}.\label{QBBtext}
\end{equation}
In the case of Gaussian states, we can compute these quantities by means of
closed analytical formulas~\cite{pirandola2008computable}.

We can write the $s$-overlap as~\cite{pirandola2008computable}%
\begin{equation}
C_{s}=2^{N}\sqrt{\frac{\det\mathbf{\Pi}_{s}}{\det\mathbf{\Sigma}_{s}}}%
\exp\left(  -\frac{\mathbf{d}^{T}\mathbf{\Sigma}_{s}^{-1}\mathbf{d}}%
{2}\right)  ,
\end{equation}
where $\mathbf{d}=\mathbf{x}_{0}-\mathbf{x}_{1}$. Here $\mathbf{\Pi}_{s}$ and
$\mathbf{\Sigma}_{s}$ are defined as
\begin{equation}
\mathbf{\Pi}_{s}:=G_{s}(\mathbf{V}_{0}^{\oplus})G_{1-s}(\mathbf{V}_{1}%
^{\oplus}),
\end{equation}
\vspace{-0.5cm}
\begin{equation}
\mathbf{\Sigma}_{s}:=\mathbf{S}_{0}\left[  \Lambda_{s}\left(  \mathbf{V}%
_{0}^{\oplus}\right)  \right]  \mathbf{S}_{0}^{T}+\mathbf{S}_{1}\left[
\Lambda_{1-s}\left(  \mathbf{V}_{1}^{\oplus}\right)  \right]  \mathbf{S}%
_{1}^{T},
\end{equation}
introducing the two real functions
\begin{align}
G_{s}(x)  &  =\frac{1}{(x+1/2)^{s}-(x-1/2)^{s}}\nonumber\\
\Lambda_{s}(x)  &  =\frac{(x+1/2)^{s}+(x-1/2)^{s}}{(x+1/2)^{s}-(x-1/2)^{s}},
\end{align}
calculated over the Williamson forms $\mathbf{V}_{i}^{\oplus}%
:=\mathbf{\bigoplus}_{k=1}^{N}\nu_{i}^{k}\mathbf{1}_{2}$, where $\mathbf{V}%
_{i}^{\oplus}\mathbf{=S}_{i}\mathbf{\mathbf{V}}_{i}^{\oplus}\mathbf{S}_{i}%
^{T}$ for symplectic $\mathbf{S}_{i}$\ and $\nu_{i}^{k}\geq1/2$ are the
symplectic spectra~\cite{serafini2003symplectic,pirandola2009correlation}.

\subsection{Quantum relative entropy for asymmetric QHT}\label{appQRE}

The quantum Stein's
lemma~\cite{hiai1991proper,ogawa2005strong} tells us that the quantum relative
entropy 
\begin{equation}
D\left(  \hat{\rho}_{0}||\hat{\rho}_{1}\right)
=\Tr [\hat{\rho}_0(\ln \hat{\rho}_0 - \ln \hat{\rho}_1)],
\end{equation}
between two quantum
states, $\hat{\rho}_{0}$ and $\hat{\rho}_{1}$, is the optimal decay rate for
the type-II error probability, given some fixed constraint, $P_{\text{fa}%
}<\epsilon$, on the type-I error
probability. Defining the quantum relative entropy variance
\begin{equation}
V\left(  \hat{\rho}_{0}||\hat{\rho}_{1}\right)
=\Tr [\hat{\rho}_0(\ln \hat{\rho}_0 - \ln \hat{\rho}_1)^2]-[D\left(  \hat
{\rho}_{0}||\hat{\rho}_{1}\right)  ]^{2},
\end{equation}
and in turn establish that the optimal type-II (missed detection) error
probability, for sample size $M$, takes the exponential
form~\cite{li2014second}
\begin{equation}
\begin{split}
P_{\text{md}}=\exp\Big\{  -\Big[  MD\left(  \hat{\rho}_{0}||\hat{\rho}%
_{1}\right) &+\sqrt{MV\left(  \hat{\rho}_{0}||\hat{\rho}_{1}\right)  }%
\Phi^{-1}(\epsilon) \\& + \mathcal{O}(\log M)\Big]  \Big\}
,\label{typeIIerrorexpform}%
\end{split}
\end{equation}
where $\epsilon\in(0,1)$ bounds $P_{\text{fa}}$\ and
\begin{equation}
\Phi(y):=\frac{1}{\sqrt{2\pi}}\int_{-\infty}^{y}dx\exp\left(  -x^{2}/2\right)
\end{equation}
is the cumulative of a normal distribution. More precisely, for finite
third-order moment (as in the present case) and sufficiently large $M$, we may
write the upper bound~\cite[Theorem~5]{li2014second}%
\begin{equation}
\begin{split}
P_{\text{md}}\leq \tilde{P}_{\text{md}}:=& \exp\Big\{  -\Big[  MD\left(
\hat{\rho}_{0}||\hat{\rho}_{1}\right)  \\&+\sqrt{MV\left(  \hat{\rho}_{0}%
||\hat{\rho}_{1}\right)  }\Phi^{-1}(\epsilon)+\mathcal{O}(1)\Big]  \Big\}
.\label{ROCub}%
\end{split}
\end{equation}

We can write explicit formulas for the relative entropy $D\left(  \hat{\rho
}_{0}||\hat{\rho}_{1}\right)  $\ and the relative entropy variance $V\left(
\hat{\rho}_{0}||\hat{\rho}_{1}\right)  $ of two arbitrary $N$-mode Gaussian
states, $\hat{\rho}_{0}(\mathbf{x}_{0},\mathbf{V}_{0})$ and $\hat{\rho}%
_{1}(\mathbf{x}_{1},\mathbf{V}_{1})$. The first one is given by~\cite{PLOB}
\begin{equation}
D\left(  \hat{\rho}_{0}||\hat{\rho}_{1}\right)  =-\Sigma\left(  \mathbf{V}%
_{0},\mathbf{V}_{0}\right)  +\Sigma\left(  \mathbf{V}_{0},\mathbf{V}%
_{1}\right)  ,
\end{equation}
where we have defined the function
\begin{equation}
\Sigma\left(  \mathbf{V}_{0},\mathbf{V}_{1}\right)  =\frac{\ln\mathrm{det}%
\left(  \mathbf{V}_{1}+\frac{i\mathbf{\Omega}}{2}\right)  +\Tr\left(
\mathbf{V}_{0}\mathbf{G}_{1}\right)  +\delta^{T}\mathbf{G}_{1}\delta}{2},
\end{equation}
with $\delta=\mathbf{x}_{0}-\mathbf{x}_{1}$ and $\mathbf{G}_{1}%
=2i\boldsymbol{\Omega}\coth^{-1}\left(  2i\mathbf{V}_{1}\boldsymbol{\Omega
}\right)  $ being the Gibbs matrix~\cite{BanchiPRL}. The second one is given
by%
\begin{equation}
V\left(  \hat{\rho}_{0}||\hat{\rho}_{1}\right)  =\frac{\Tr\left[
(\mathbf{\Gamma}\mathbf{V}_{0})^{2}\right]  }{2}+\frac{\Tr\left[
(\mathbf{\Gamma}\mathbf{\Omega})^{2}\right]  }{8}+\delta^{T}\mathbf{G}%
_{1}\mathbf{V}_{0}\mathbf{G}_{1}\delta,
\end{equation}
where $\mathbf{\Gamma}=\mathbf{G}_{0}-\mathbf{G}_{1}$~\cite{RevQKD} (see also Ref.~\cite{wilde2017qht} and Ref.~\cite[App.~A]{LaurenzaBounds}).

\end{document}